\documentclass[twocolumn,A4]{article}

\usepackage[dvips]{graphics}

\usepackage{setspace}
\usepackage{color}
\definecolor{gold}{rgb}{0.85,0.66,0}
\definecolor{dblue}{rgb}{0,0,0.8}

\begin{document}

\onecolumn

\begin{center}
{\bf{\Large {\textcolor{gold}{Quantum Transport through Organic 
Molecules}}}}\\
~\\
{\textcolor{dblue}{Santanu K. Maiti}}$^{1,2,*}$ \\
~\\
{\em $^1$Theoretical Condensed Matter Physics Division,
Saha Institute of Nuclear Physics, \\
1/AF, Bidhannagar, Kolkata-700 064, India \\
$^2$Department of Physics, Narasinha Dutt College,
129 Belilious Road, Howrah-711 101, India} \\
~\\
{\bf Abstract}
\end{center}
We explore electron transport properties for the model of benzene-$1$, 
$4$-dithiolate (BDT) molecule and for some other geometric models of 
benzene molecule attached to two semi-infinite one-dimensional metallic 
electrodes using the Green's function formalism. An analytic approach, 
based on a simple tight-binding framework, is presented to describe 
electron transport through the molecular wires. Electronic transport in 
such molecular systems is strongly affected by the geometry of the 
molecules as well as their coupling to the side-attached electrodes. 
Conductance reveals resonant peaks associated with the molecular energy 
eigenstates providing several complex spectra. Current passing through 
the molecules shows staircase-like behavior with sharp steps in the weak 
molecule-to-electrode coupling limit, while it varies quite continuously 
with the applied bias voltage in the limit of strong molecular coupling. 
In the presence of transverse magnetic field, conductance exhibits 
oscillatory behavior with flux $\phi$, threaded by the molecular ring, 
showing $\phi_0$ ($=ch/e$) flux-quantum periodicity. Though, conductance 
changes in the presence of transverse magnetic field, but the 
current-voltage characteristics are not significantly affected by 
this field.

\vskip 1cm
\begin{flushleft}
\end{flushleft}

\vskip 2.7in
\noindent
{\bf ~$^*$Corresponding Author}: Santanu K. Maiti

Electronic mail: santanu.maiti@saha.ac.in

\newpage

\twocolumn

\section{{\textsl{Introduction}}}

Molecular electronics is an essential technological concept of 
fast-growing interest since molecules constitute promising building 
blocks for future generation of electronic devices. Understanding of 
the fundamental processes of electron conduction through individual 
molecules is a most important requirement for the purposeful design 
of molecules for electronic functionalities. Electron transport 
through molecules was first studied theoretically in $1974$~\cite{aviram}. 
Following this work, several experiments~\cite{metz,reed1,fish} have 
been performed through molecules placed between two electrodes with 
few nanometer separation. Transport properties in such molecular 
systems cannot be studied by using the conventional procedure done 
in electronics~\cite{fisc} i.e., by solving the Boltzmann's equation. 
Full quantum mechanical treatment is required to characterize the 
transport through molecules. The operation of such two-terminal devices 
is due to an applied bias. Current passing across the junction is 
strongly nonlinear function of applied bias voltage and its detailed 
description is quite complex. Transport properties of these systems 
are associated with many quantum effects, like as quantization of 
energy levels and quantum interference effects~\cite{baer1,baer2,baer3,
walc1,walc2,san1,san2,san3,san4,tagami} of electron 
waves. A quantitative understanding of physical mechanisms underlying 
the operation of nanoscale devices remains a major challenge in
nanoelectronics research. Here, we focus on the molecular transport 
that are currently the subject of substantial experimental, theoretical 
and technological interest. These molecular systems can act as gates, 
switches, or transport elements, providing new molecular functions 
that need to be well characterized and understood. In an experiment 
Reed {\em et al.}~\cite{reed2} have investigated the conductance and 
current-voltage characteristics of benzene-$1$, $4$-dithiolate molecule 
in a two terminal geometry which is highly reproducible. This motivates 
us to calculate electron transport properties through benzene molecules 
attached to two semi-infinite metallic electrodes. 
  
In the present work, we describe theoretically the electronic transport 
properties for the model of benzene-$1$, $4$-dithiolate (BDT) molecule 
and also for different other geometric models of benzene molecule, 
attached to two metallic electrodes, using the Green's function 
technique. The transport properties strongly 
depend on the geometry of the molecule and the molecular coupling 
strength to the side-attached electrodes. Based on the scanning probe 
technique measurement, conductance of such molecular systems is directly 
measured~\cite{hong,tans,bock,don,cui,schon}. Theoretically there 
exist several formulations~\cite{mol,roth} for the calculation of 
conductance by using the Landauer conductance formula and the seminal 
$1974$ paper of Aviram and Ratner~\cite{aviram}. At much low 
temperatures and bias voltages, electron transport becomes coherent 
through the molecule. Here we assume that the dissipation and 
equilibration processes occur only in the two contacting electrodes 
and this approximation enables to describe the propagation of an 
electron by means of single particle Green's function. This theory 
is much more flexible than any other theoretical approach and also 
applicable to any system described by a Hamiltonian with a localized 
orbital basis. By using this method, electronic transport of any 
system can be studied very easily with a few computational cost. In 
that case we have to know only the Hamiltonian matrix for the molecule 
without knowing anything about the electronic wave functions. 

Here, we reproduce an analytic approach based on a simple tight-binding
framework to investigate electron transport for the molecules taken into 
consideration. Several {\em ab initio} methods are available for the
calculation of conductance~\cite{yal,ven,xue,tay,der,dam}, yet simple 
parametric approaches~\cite{baer1,baer2,baer3,walc1,walc2,san1,san2,
san3,san4,muj1,muj2,sam,hjo} are much more useful. The parametric study is 
motivated by the fact that the {\em ab initio} theories are 
computationally too expensive and here we do attention only on the 
qualitative effects rather than the quantitative ones. 

We organize the paper as follows. Following the brief introduction 
(Section $1$), in Section $2$, we describe the formulation of conductance 
$g$ by calculating the transmission probability $T$ and current $I$ for 
any finite size conducting system attached to two semi-infinite 
one-dimensional ($1$D) metallic electrodes using the Green's function 
technique. Section $3$ presents the significant results, where 
sub-section $3.1$ focuses the results for single benzene molecules 
with different geometries, while in sub-section $3.2$, we concentrate on 
the results for an array of benzene molecules. The effect of transverse 
magnetic field on electron transport through a single benzene molecule 
is explored in sub-section $3.3$. Finally, the summary of our results is 
available in Section $4$.

\section{{\textsl{A brief description of the theoretical formulation}}}

Here we give a brief description for the calculation of transmission 
probability ($T$), conductance ($g$) and current ($I$) through a finite 
size conducting system attached to two semi-infinite $1$D electrodes 
within the Green's function approach.

Let us consider a $1$D conductor with $N$ number of sites (filled black
circles) connected to two semi-infinite $1$D electrodes, viz, source and
\begin{figure}[ht]
{\centering \resizebox*{7.5cm}{2.1cm}{\includegraphics{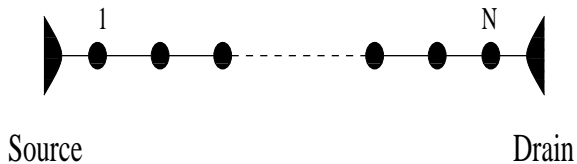}}\par}
\caption{{\textsl{Schematic view of a one-dimensional conductor with 
$N$ number of sites (filled black circles) attached with two electrodes. 
The first and the last sites are labeled by $1$ and $N$, respectively.}}}
\label{dot}
\end{figure}
drain, as shown in Fig.~\ref{dot}. The conducting system within the 
electrodes can be anything like an array of few quantum dots, or a 
single molecule, or an array of few molecules, etc. At much low 
temperatures and bias voltage, the conductance $g$ of the conductor 
can be written by using the Landauer conductance formula as~\cite{datta},
\begin{equation}
g=\frac{2e^2}{h}T
\label{land}
\end{equation}
where, $T$ is the transmission probability of an electron through the 
conductor. It ($T$) can be expressed in terms of the Green's function 
of the conductor and the coupling of the conductor to the electrodes 
through the expression~\cite{datta},
\begin{equation}
T={\mbox{Tr}} \left[\Gamma_S G_c^r \Gamma_D G_c^a\right]
\label{trans1}
\end{equation}
where, $G_c^r$ and $G_c^a$ are the retarded and advanced Green's functions 
of the conductor, respectively. $\Gamma_S$ and $\Gamma_D$ are the coupling 
terms of the conductor due to the coupling to the source and drain, 
respectively. For the complete system i.e., the conductor and two 
electrodes, the Green's function is defined as,
\begin{equation}
G=\left(E-H\right)^{-1}
\end{equation}
where, $E$ is the injecting energy of the source electron. This Green's 
function corresponds to the inversion of an infinite matrix which 
consists of the finite conductor and two semi-infinite electrodes. It 
can be partitioned into different sub-matrices that correspond to the 
individual sub-systems.

The effective Green's function for the conductor can be written in 
the form,
\begin{equation}
G_c=\left(E-H_c-\Sigma_S-\Sigma_D\right)^{-1}
\end{equation}
where, $H_c$ is the tight-binding Hamiltonian of the conductor. Withing the
non-interacting electron picture, the Hamiltonian for the conductor 
becomes,
\begin{equation}
H_c=\sum_i \epsilon_i c_i^{\dagger} c_i + \sum_{<ij>}t 
\left(c_i^{\dagger}c_j + c_j^{\dagger}c_i \right)
\label{hamil1}
\end{equation}
where, $\epsilon_i$ is the on-site energy and $t$ is the nearest-neighbor 
hopping integral. Here, $\Sigma_S=h_{Sc}^{\dagger} g_S h_{Sc}$ and 
$\Sigma_D=h_{Dc} g_D h_{Dc}^{\dagger}$ are the self-energy terms due to 
the two electrodes. $g_S$ and $g_D$ correspond to the Green's functions 
for the source and drain, respectively. $h_{Sc}$ and $h_{Dc}$ are the 
coupling matrices and they are non-zero only for the adjacent points 
in the conductor, $1$ and $N$ as shown in Fig.~\ref{dot}, and the 
electrodes, respectively. The coupling terms $\Gamma_S$ and $\Gamma_D$ 
for the conductor can be calculated through the relation~\cite{datta},
\begin{equation}
\Gamma_{\{S,D\}}=i\left[\Sigma_{\{S,D\}}^r-\Sigma_{\{S,D\}}^a\right]
\end{equation}
where, $\Sigma_{\{S,D\}}^r$ and $\Sigma_{\{S,D\}}^a$ are the retarded and
advanced self-energies, respectively, and they are conjugate to each
other. Datta {\em et al.}~\cite{tian} have shown that the self-energies 
can be expressed in the form,
\begin{equation}
\Sigma_{\{S,D\}}^r=\Lambda_{\{S,D\}}-i \Delta_{\{S,D\}}
\end{equation} 
where, $\Lambda_{\{S,D\}}$ are the real parts of the self-energies which 
correspond to the shift of the energy eigenstates of the conductor and the 
imaginary parts $\Delta_{\{S,D\}}$ of the self-energies represent the 
broadening of these energy levels. Since this broadening is much higher 
than the thermal broadening, we restrict our all calculations only 
at absolute zero temperature. Thus, the coupling terms $\Gamma_S$ and 
$\Gamma_D$ can be written in terms of the retarded self-energy 
as~\cite{datta},
\begin{equation}
\Gamma_{\{S,D\}}=-2 {\mbox{Im}} \left[\Sigma_{\{S,D\}}^r\right]
\end{equation} 
All the information regarding the conductor-to-electrode coupling are 
included into the two self energies and are analyzed through the use of 
Newns-Anderson chemisorption theory~\cite{muj1,muj2}. The detailed 
description of this theory is available in these two references.

Calculating the self-energies, the coupling terms $\Gamma_S$ and 
$\Gamma_D$ can be easily obtained and then the transmission probability 
($T$) will be evaluated from the expression given in Eq.~\ref{trans1}.
 
As the coupling matrices $h_{Sc}$ and $h_{Dc}$ are non-zero only for the 
adjacent points of the conductor, $1$ and $N$ as shown in Fig.~\ref{dot},
the transmission probability becomes~\cite{datta},
\begin{equation}
T(E)=4~\Delta_{11}^S(E) ~\Delta_{NN}^D(E) ~|G_{1N}(E)|^2
\label{trans2}
\end{equation}

The current passing through the conductor is depicted as a single-electron 
scattering process between the two reservoirs of charge carriers and the 
current-voltage ($I$-$V$) relation is evaluated through the 
expression~\cite{datta},
\begin{equation}
I(V)=\frac{2e}{h}\int \limits_{E_F-eV/2}^{E_F+eV/2} T(E) ~dE
\end{equation}
where $E_F$ is the equilibrium Fermi energy. For the sake of simplicity, 
here we assume that the entire voltage is dropped across the 
conductor-electrode interfaces and this assumption does not affect 
significantly the $I$-$V$ characteristics. With the expression of 
$T(E)$ given in Eq.~\ref{trans2}, the final form of $I(V)$ becomes, 
\begin{eqnarray}
I(V) &=& \frac{8e}{h}\int \limits_{E_F-eV/2}^{E_F+eV/2}
\Delta_{11}^S(E) ~\Delta_{NN}^D(E) \nonumber \\
& & ~\times |G_{1N}(E)|^2 ~dE
\label{curr}
\end{eqnarray} 
Eqs.~\ref{land}, \ref{trans2} and \ref{curr} are the final working 
formule for the calculation of conductance $g$ and $I$-$V$ 
characteristics, respectively, for any finite size conductor 
sandwiched between two electrodes. 

Using the above formulation, in the forthcoming sections we shall
describe the behavior of electron transport for some specific models 
of benzene molecules. For simplicity, we take the unit $c=h=e=1$
in our present calculations. 

\section{{\textsl{Results and discussion}}}

\subsection{{\textsl{Single benzene molecules attached to electrodes}}} 

\begin{figure}[ht]
{\centering \resizebox*{6cm}{8cm}{\includegraphics{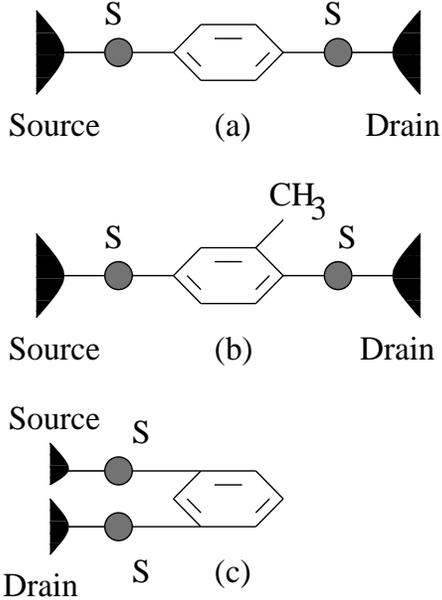}}\par}
\caption{{\textsl{Schematic view of single benzene molecules attached 
to the electrodes via thiol (SH bond) groups, where (a) benzene molecule
attached symmetrically, (b) symmetry is broken by adding the chemical 
substituent $CH_3$ in upper arm of the molecular ring and (c) chemical 
substituent free benzene molecule attached asymmetrically.}}} 
\label{benzene1}
\end{figure}

In this section we study quantum transport through single benzene 
molecules and investigate the geometrical effect on conductance-energy 
($g$-$E$) and current-voltage ($I$-$V$) characteristics. The schematic 
representations of the single benzene molecules attached to the two 
electrodes via thiol (SH bond) groups are shown in Fig.~\ref{benzene1}. 
In experiments, two electrodes made from gold (Au) are used and 
molecules coupled to the electrodes by thiol groups in the 
chemisorption technique where hydrogen (H) atoms remove and sulfur (S) 
atoms reside. In Fig.~\ref{benzene1}(a), thiol groups are attached 
symmetrically at $1$ and $4$ positions of the benzene molecule and 
it is the so-called benzene-$1$, $4$-dithiolate molecule. The symmetry 
can be broken by adding a chemical substituent group at any one arm 
of the molecular ring 
(Fig.~\ref{benzene1}(b)) or by contacting the electrodes asymmetrically
as given in Fig.~\ref{benzene1}(c), where the electrodes are connected at
$1$ and $5$ positions (according to the clockwise direction). The idea
of considering such different geometries is that only in this way the 
interference conditions are changed and they have strong influence on the 
electron transport through molecular bridges. 

All the essential features of electron transport are discussed in the
two distinct regimes. One is defined as $\tau_{\{S,D\}} << t$, called 
the weak-coupling regime and the other one is mentioned as $\tau_{\{S,D\}} 
\sim t$, called the strong-coupling regime, where $\tau_S$ and $\tau_D$ 
are the hopping strengths of the molecule to the source and drain, 
\begin{figure}[ht]
{\centering \resizebox*{7.5cm}{11.5cm}
{\includegraphics{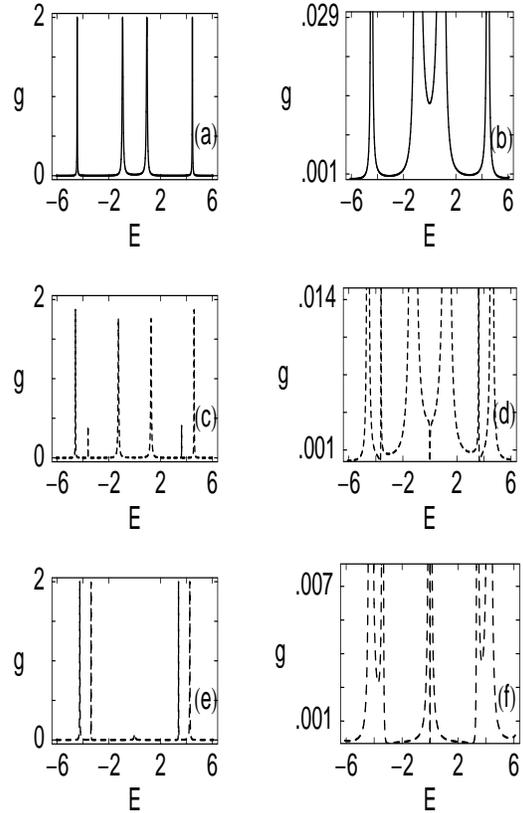}}\par}
\caption{{\textsl{$g$-$E$ spectra for the benzene molecules in the 
weak-coupling limit. (a), (c) and (e) correspond to the results for 
the bridges given in Figs.~\ref{benzene1}(a), (b) and (c), respectively. 
Lower parts of the curves plotted in the $1$st column are redrawn in the 
$2$nd column to explain the conductance behavior much more clearly.}}}
\label{benzcondlow}
\end{figure}
respectively. Throughout the work, common set of values of the parameters 
used in our calculations for these two limiting cases are as follows: 
$\tau_S=\tau_D=0.5$, $t=3$ (weak-coupling) and $\tau_S=\tau_D=2.5$,
$t=3$ (strong-coupling). The Fermi energy $E_F$ is fixed at $0$.

Figure~\ref{benzcondlow} shows the variation of conductance $g$ as 
a function of injecting electron energy $E$ for the single benzene 
molecules in the limit of weak-coupling, where (a), (c) and (e) 
correspond to the results for the models given in Figs.~\ref{benzene1}(a), 
(b) and (c), respectively. To observe different resonant and 
anti-resonant peaks much more clearly, in the $2$nd column of 
Fig.~\ref{benzcondlow} we plot the lower portions of the curves given 
in the $1$st column of this figure. From the curves plotted in the 
$1st$ column, it is observed that the conductance shows sharp resonant 
peaks for some particular energy values, while it vanishes for all 
other energies. At resonance, conductance 
\begin{figure}[ht]
{\centering \resizebox*{7.5cm}{5cm}
{\includegraphics{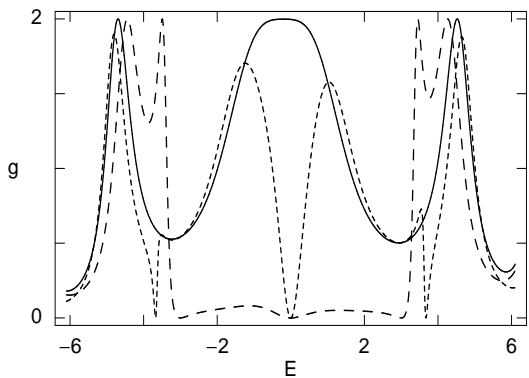}}\par}
\caption{{\textsl{$g$-$E$ curves for the benzene molecules in the limit 
of strong-coupling. The solid, dotted and dashed curves correspond to 
the results for the models given in Figs.~\ref{benzene1}(a), (b) and (c),
respectively.}}} 
\label{benzcondhigh}
\end{figure}
approaches to $2$, and accordingly, the transmission probability $T$ 
becomes unity since we get the relation $g=2T$ from the Landauer 
conductance formula (see Eq.~\ref{land} with $e=h=1$ in our present 
formulation). The resonant peaks in the conductance spectrum coincide 
with the eigenenergies of the single benzene molecules. Thus we can say
that the conductance spectrum manifests itself the electronic structure 
of the molecules.

The effect of quantum interference becomes much more clear from the
results plotted in Figs.~\ref{benzcondlow}(c) and (e). These results 
predict that some of the conductance peaks do not reach to unity 
anymore and get much reduced amplitude. This feature can be understood 
as follows. Electrons are transmitted through the molecule from source
to drain. Electronic waves propagating along the two arms of the 
molecular ring acquire a phase shift among themselves, and therefore, 
the probability amplitude of getting an electron after traversing 
through the molecule becomes higher or smaller. It manifests itself 
especially as transmittance cancellations and provides anti-resonant
states in the conductance spectrum. Thus, we can predict that the 
electronic transport is significantly influenced by the quantum 
interference effect i.e., the molecule- electrode interface structure.

The effect of molecular coupling has a significant role in electron
transport. In the strong molecule-to-electrode coupling limit, all the
resonant peaks get substantial widths, as shown in Fig.~\ref{benzcondhigh},
where the solid, dotted and dashed curves correspond to the results for 
the molecular bridges presented in Figs.~\ref{benzene1}(a), (b) and (c), 
respectively. The enhancement of the resonance widths is due to the 
broadening of the molecular energy levels in the limit of strong 
molecular coupling, where the contribution comes from the imaginary 
parts of the self-energies $\Sigma_S$ and $\Sigma_D$~\cite{datta}.

The other significant feature observed from the conductance spectrum 
is the appearance of anti-resonant states where the conductance vanishes
exactly to zero. Such anti-resonant states are specific to the 
interferometric nature of the scattering and they do not appear
in traditional one-dimensional scattering problems. It is also noticed 
that the positions of the anti-resonances on energy scale are independent 
of the molecule-to-electrode coupling strength which are clearly observed 
from the results plotted in the second column of Figs.~\ref{benzcondlow} 
and \ref{benzcondhigh}. Since the widths of these states are very small 
they do not provide any significant contribution in the current-voltage 
characteristics. However, the variations of interference conditions have 
strong influence on the magnitude of current flowing through the
molecular bridges.

\begin{figure}[ht]
{\centering \resizebox*{7.5cm}{10cm}{\includegraphics{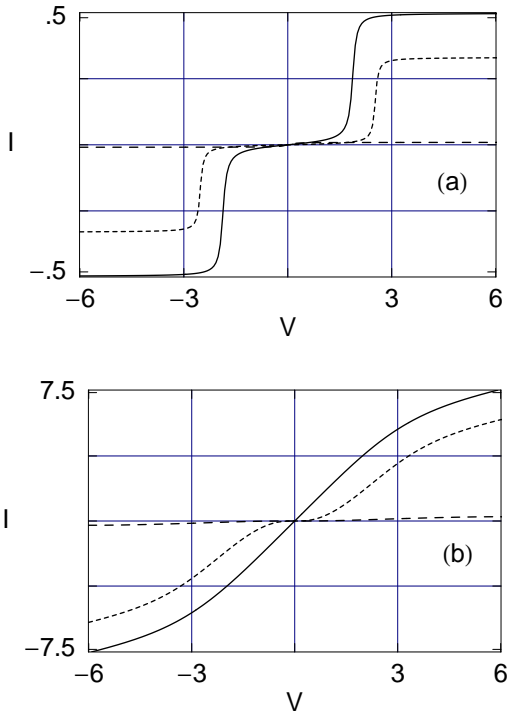}}\par}
\caption{{\textsl{Current $I$ as a function of applied bias voltage $V$ 
for the single benzene molecules, where the solid, dotted and dashed 
lines correspond to the results for the molecular bridges given in 
Figs.~\ref{benzene1}(a), (b) and (c), respectively. (a) weak-coupling 
(b) strong-coupling.}}}
\label{benzcurr}
\end{figure}

The scenario of electron transfer through the molecular wires becomes much
more clearly observed from the current-voltage ($I$-$V$) spectra. Current
through the molecular system is computed by the integration method of
the transmission function $T$. The feature of the transmission function
is exactly similar to that of the conductance spectrum, differ only in
magnitude by a factor $2$, according to the Landauer conductance formula
$g=2T$. In Fig.~\ref{benzcurr}, we display the current-voltage 
characteristics for the molecular bridges shown in Fig.~\ref{benzene1},
where (a) and (b) represent the currents for the weak- and strong-coupling 
limits, respectively. The solid, dotted and dashed lines correspond to
the results for the molecular bridges given in Figs.~\ref{benzene1}(a), 
(b) and (c), respectively. It is observed that, in the limit of weak 
molecular coupling current shows staircase-like structure with sharp 
steps. This is due to the existence of discrete molecular resonances 
as shown in Fig.~\ref{benzcondlow}. With the increase in applied bias 
voltage $V$, the difference in electrochemical potentials between the 
two electrodes $(\mu_1 - \mu_2)$ increases, favoring more number of 
discrete energy levels to fall in that range, and therefore, more energy 
channels are accessible to the injected electrons to pass through the 
molecule from the source to drain. Incorporation of a single discrete 
energy level i.e., a discrete quantized conduction channel, between the 
range $(\mu_1 - \mu_2)$ provides a jump in the $I$-$V$ characteristics.
The shape and height of these current steps depend on the width of the 
molecular resonances. With the increase of molecule-to-electrode coupling 
strength, current varies continuously as a function of the applied bias 
voltage and gets much higher values, as shown in Fig.~\ref{benzcurr}(b). 
Both for the weak- and strong-coupling limits, current amplitude gets 
reduced for the asymmetric bridges (see the dotted and dashed curves), 
and for the bridge given in Fig.~\ref{benzene1}(c) current almost vanishes. 
This is solely due to the effect of quantum interference between the
electronic waves passing through different arms of the molecular ring.
Hence, it can be clearly emphasized that, designing a molecular device 
is strongly influenced by the molecule-to-electrode interface structure
as well as molecular coupling strength.

\subsection{{\textsl{Array of benzene molecules attached to electrodes}}}

\begin{figure}[ht]
{\centering \resizebox*{7.5cm}{9cm}{\includegraphics{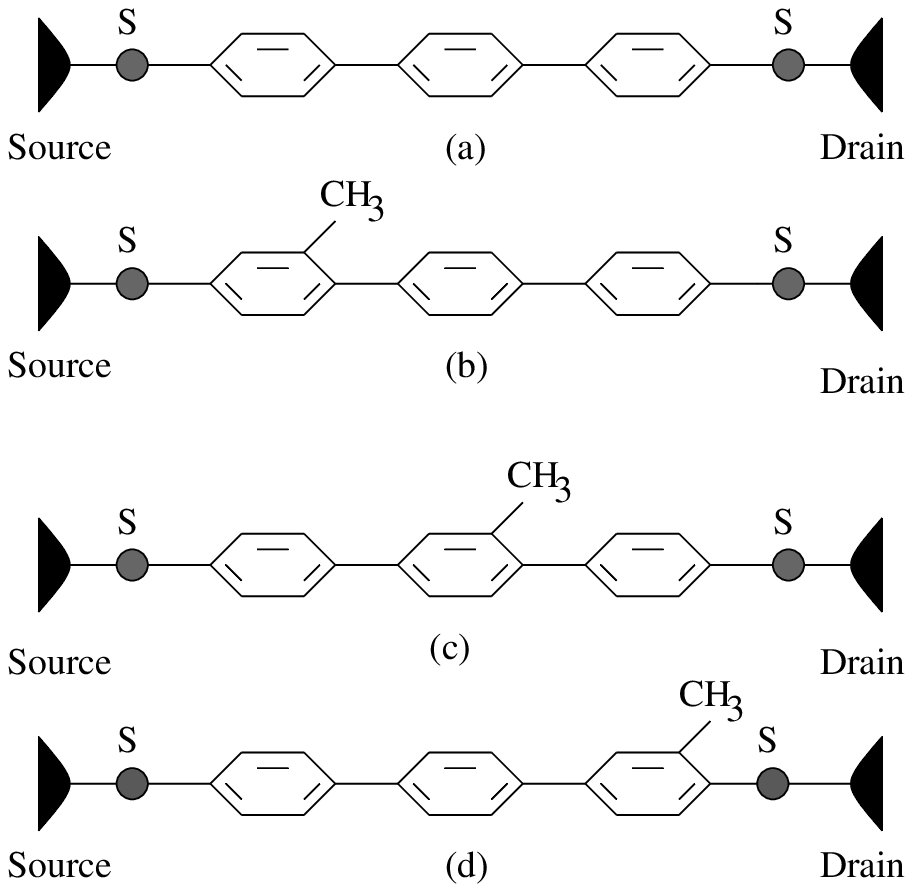}}\par}
\caption{{\textsl{Array of benzene molecules attached to the electrodes, 
source and drain, via thiol (SH bond) groups, where (a) without any 
chemical substituent, (b) chemical substituent ($CH_3$) at $1$st molecule, 
(c) chemical substituent ($CH_3$) at $2$nd molecule and (d) chemical 
substituent ($CH_3$) at $3$rd molecule.}}}
\label{benzene2}
\end{figure}

This section follows electron transport through an array of benzene 
molecules. Schematic representations for some array of benzene molecules 
are shown in Fig.~\ref{benzene2}, where the molecules are attached to the
electrodes via thiol groups. Figure~\ref{benzene2}(a) represents the array
of benzene molecules without any chemical substituent, while for the other
three arrays the chemical substituent $CH_3$ is added at different 
molecules as shown in Figs.~\ref{benzene2}(b), (c) and (d), respectively. 

For these molecular bridges we also get the resonant and anti-resonant 
peaks (not plotted here) as a function of energy $E$ similar to the cases 
of single molecular bridge systems. Since the resonant peaks are 
associated with the eigenenergies of the molecule itself, here we get 
more resonant peaks compared to Figs.~\ref{benzcondlow} and 
\ref{benzcondhigh}. Due to the breaking of molecular symmetry by adding 
the chemical substituent group $CH_3$, more anti-resonant peaks appear 
for these molecular wires. In the weak-coupling limit, resonant peaks are 
very sharp, while in the strong molecular coupling they get broadened 
substantially.

\begin{figure}[ht]
{\centering \resizebox*{7.5cm}{9cm}{\includegraphics{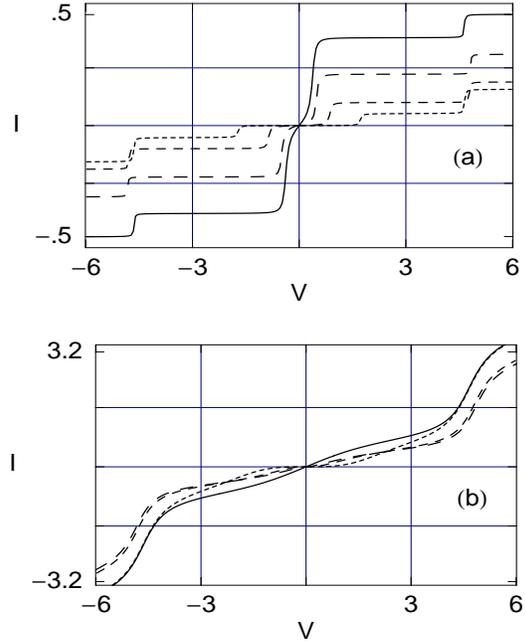}}\par}
\caption{{\textsl{Current $I$ as a function of applied bias voltage $V$ 
for the array of benzene molecules, where the solid, dotted, small dashed 
and dashed lines correspond to the results for the molecular bridges shown 
in Figs.~\ref{benzene2}(a), (b), (c) and (d), respectively. 
(a) weak-coupling limit and (b) strong-coupling limit.}}}
\label{benzarraycurr}
\end{figure}

The scenario of the current-voltage characteristics for these array of 
molecules is presented in Fig.~\ref{benzarraycurr}, where (a) and (b)
correspond to the results in the limits of weak- and strong-coupling,
respectively. The solid, dotted, small dashed and dashed curves are, 
respectively, for the bridges presented in Figs.~\ref{benzene2}(a), 
(b), (c) and (d). Current shows sharp staircase-like behavior (see 
Fig.~\ref{benzarraycurr}(a)) in the limit of weak-coupling associated 
with the discrete molecular energy levels. Here, more steps in the 
$I$-$V$ curve appears than the single molecular bridge systems (see 
Fig.~\ref{benzcurr}(a)). With the increase of molecular coupling, 
current varies almost continuously as given in Fig.~\ref{benzarraycurr}(b) 
and achieves much higher values than the current observed in the weak 
molecular coupling. Finally, here we also like to mention that the current
amplitude gets reduced with the addition of chemical substituent group 
$CH_3$ irrespective of its position in the array.

\subsection{{\textsl{Single benzene molecules attached to electrodes 
in presence of magnetic flux $\phi$}}}

At the end, here we study the behavior of conductance $g$ of the single 
benzene molecules given in Fig.~\ref{benzene1} in the presence of 
transverse magnetic field $B$. We assume that the transverse magnetic 
field passes through the molecular ring in such a way that it doesn't 
penetrate the circumference of the ring 
\begin{figure}[ht]
{\centering \resizebox*{7.5cm}{11.5cm}{\includegraphics{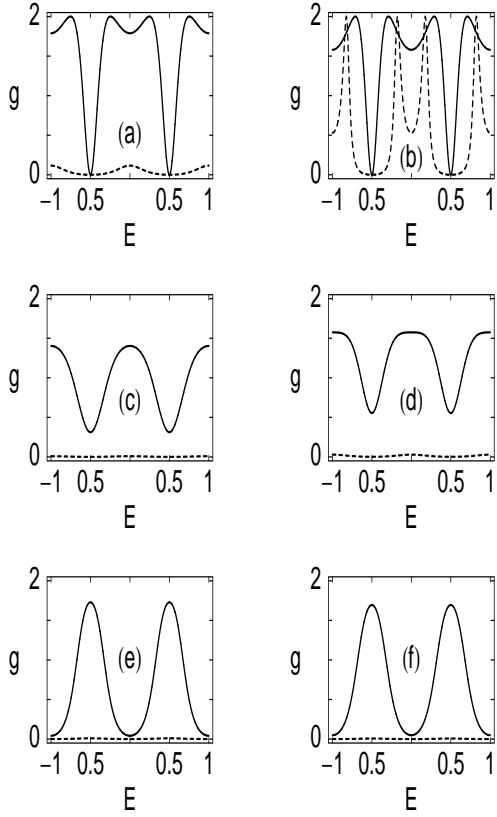}}\par}
\caption{{\textsl{$g$-$\phi$ characteristics of single benzene molecules. 
The $1$st, $2$nd and $3$rd rows represent the results for the molecular 
bridges given in Figs.~\ref{benzene1}(a), (b) and (c), respectively. 
The $1$st column corresponds to the energy $E=0.75$, while the $2$nd 
column represents the energy $E=1.0$. The solid and dotted lines 
correspond to the strong- and weak-coupling cases, respectively.}}}
\label{benzphi}
\end{figure}
anywhere, and therefore, we neglect additional Zeeman term in our 
calculations. Due to the magnetic flux $\phi$, associated with transverse
magnetic field $B$, an additional phase difference appears between the 
electronic waves traversing through different arms of the molecular ring,
and accordingly, the tight-binding Hamiltonian (Eq.~\ref{hamil1}) gets 
modified by a phase factor. The single band tight-binding Hamiltonian 
that describes the molecule in the presence of a magnetic flux $\phi$ 
can be written within the non-interacting electron picture in the 
following form,
\begin{equation}
H_c=\sum_i \epsilon_i c_i^{\dagger} c_i + \sum_{<ij>}t 
\left(e^{i\theta} c_i^{\dagger}c_j + e^{-i\theta} c_j^{\dagger}c_i 
\right)
\label{hamil2}
\end{equation}
where, $\theta=2\pi \phi/N$ represents the phase factor due to the flux 
$\phi$ threaded by the molecular ring and other symbols carry their usual 
meaning as in Eq.~\ref{hamil1}. Introducing a magnetic flux, interference 
conditions can also be changed. Both constructive and destructive 
interferences take place, and therefore, the oscillating behavior of
conductance with $\phi$ is observed. In Fig.~\ref{benzphi}, we plot 
$g$-$\phi$ characteristics of single benzene molecules, where the $1$st, 
$2$nd and $3$rd rows represent the results for the molecular models 
shown in Figs.~\ref{benzene1}(a), (b) and (c), respectively. The $1$st 
column of Fig.~\ref{benzphi} corresponds to the variation for the typical
energy $E=0.75$, while the $2$nd column of this figure represents the 
variation when the energy $E$ is fixed at $1.0$. The solid and dotted 
curves denote the results for the strong and weak molecule-to-electrode 
coupling limits, respectively. Conductance shows oscillatory behavior 
with $\phi$ exhibiting $\phi_0$ flux-quantum periodicity with extremas at
half-integer flux quantum ($\phi_0/2$ i.e., $0.5$ in our chosen unit) 
irrespective of the molecular coupling strength. In the weak-coupling 
limit, conductance almost vanishes (dotted curves) for all these three 
molecular bridges, while for the limit of strong-coupling they get much 
higher values (solid curves). Though in the presence of $\phi$ 
conductance ($g$) gets modified significantly compared to that in the
absence of any $\phi$, yet the current-voltage characteristics do not 
change appreciably. Thus, it can be emphasized that by applying such 
magnetic field, current cannot be modified significantly in these molecular 
bridge systems.

All the above pictures remain valid if the electron-electron interaction is
taken into account. The main effect of the electron correlation is to shift
and to split the resonant positions. This is due to the fact that the on-site
Coulomb repulsive energy $U$ gives a renormalization of the site energies.
Depending on the strength of the nearest-neighbor hopping integral ($t$)
compared to the on-site Coulomb interaction ($U$) different regimes appear.
For the case $t/U<<1$, the resonances and anti-resonances would split into 
two distinct narrow bands separated by the on-site Coulomb energy. On the 
other hand, for the case where $t/U>>1$, the resonances and anti-resonances 
would occur in pairs.

\section{{\textsl{Concluding remarks}}}

To conclude, we have introduced a parametric approach based on the 
tight-binding model to investigate the transport properties through 
benzene-$1$, $4$-dithiolate molecular model and some other geometric 
models of benzene molecule. Molecular geometry and molecule-to-electrode 
coupling strength have significant roles on the electronic transport 
through such molecular wires. Conductance shows resonant peaks for 
some particular energies associated with the molecular energy levels, 
while in all other cases it drops to zero. Due to breaking of the 
molecular symmetry more anti-resonant peaks appear in the conductance 
spectra and their positions are independent of the molecular coupling 
strength.

In the weak-coupling limit, current shows staircase-like structure, while 
it gets a continuous variation with the applied bias voltage $V$ in the 
limit of strong-coupling. For a fixed molecular coupling, current amplitude
in symmetric molecule is quite higher than the asymmetric one, which is
controlled by the quantum interference effect of the electronic waves 
passing through different arms of the molecular ring. It is also observed
that for a particular molecular wire, current amplitude can be enhanced
an order of magnitude by changing the molecular coupling strength.

Lastly, we have studied the behavior of conductance for the single benzene
molecules in presence of the transverse magnetic field. Conductance shows 
oscillatory behavior with flux $\phi$ and gives $\phi_0$ flux-quantum 
periodicity. We have observed that the conductance changes in presence of 
magnetic field but the current-voltage characteristics remain same and thus 
we can predict that the current cannot be controlled significantly by means 
of such magnetic field in these molecular bridges.

In the present paper we have done all the calculations by ignoring
the effects of the temperature, electron-electron correlation, etc.
Due to these factors, any scattering process that appears in the
molecular ring would have influence on electronic phases, and, in
consequences can disturb the quantum interference effects. Here we
have assumed that, in our sample all these effects are too small, and
accordingly, we have neglected all these factors in this particular
study.

\vskip 0.3in
\noindent
{\bf\Large \textsl{Acknowledgments}}
\vskip 0.2in
\noindent
It is a pleasure to thank Atikur Rahman (my best friend) and Prof. S. N.
Karmakar for many helpful comments and suggestions.

\end{document}